# Propagation of Nonlinear Acoustic Waves in the Suspension of Ultrasound Contrast Agents Part I: Equation for Counting Resonance Effects and Revealing Acoustic Localization

Lang Xia

Email: langxia.org@gmail.com

**Abstract**

The oscillations of ultrasound contrast agents are of particular importance to the understanding of the propagation of acoustic waves in the bubbly liquids (suspensions of ultrasound contrast agents). Acoustic waves propagating in bubbly liquids have been investigated extensively. Little has been dedicated to the resonance effects of the microbubbles on the propagating waves. Here a nonlinear partial differential equation for describing one-dimensional acoustic waves propagating near the resonance frequency of the microbubbles in bubbly liquids is obtained. The present equation recovers classical results for propagating acoustic waves with finite amplitudes in liquids and interprets the acoustic localization in bubbly liquids explicitly.

## 1. Introduction

Acoustic waves propagating through liquids containing gas bubbles (bubbly liquids) have been explored both theoretically and numerically for a long time (Foldy 1945, Wijngaarden 1968, Wijngaarden 1972, Nigmatulin, Khabeev et al. 1988, Commander and Prosperetti 1989, Ruffa 1992, Nakoryakov, Kashinsky et al. 1996, Smereka 2002, Vanhille and Campos-Pozuelo 2009, Bodunova, Konoplev et al. 2011, Nigmatulin, Gubaidullin et al. 2013). These investigations are mainly divided into two categories: Foldy treated the wave propagation in bubbly liquids as a multiple scattering problem and found that the coherent pressure satisfied the linear wave equation; whereas Wijngaarden's model treated the bubbly liquid as a continuum medium analog to the gas dynamics of a single-phase medium. Both of the above models have been widely recognized. Numerical results generated by the two models have also shown similarities in the linear region approximation (Wijngaarden 1972).

Propagations of nonlinear acoustic waves in single phase media are well described by the Kuznetsov equation (Kuznetsov 1971). For bubbly liquids, however, unified model equations for describing the nonlinear propagation is still under developing. The Kuznetsov equation may not be easily applied to bubbly liquids due to the dispersive nature of the media. Equations describing weakly nonlinear waves and weak dispersion in bubbly liquids have been discussed before (Crighton 1991). A widely spread equation was proposed by Wijngaarden for finite-amplitude waves propagating in bubbly liquids (Wijngaarden 1968, Commander and Prosperetti 1989), in which the pressure is a volume-averaged quantity. Other authors used a similar technique to derived various model equations (Nigmatulin, Khabeev et al. 1988, Lauterborn, Kurz et al. 2014).

The nonlinearity of a bubbly liquid could be substantially high when the containing bubbles are oscillating at resonance (Xia and Sarkar 2017). Little study has been found studying the resonance effects on the propagation of nonlinear acoustic waves in bubbly liquids. Here we follow the Wijngaarden's approach and derive simple nonlinear wave equations addressing resonance concerns.

## 2. Governing Equations

We treat the suspension of ultrasound contrast agents as a homogeneous medium, using average quantities to derive the nonlinear wave equations in the bubbly liquid.

### 2.1. Wave Equations

By assuming the bubbles in a bubbly liquid to be far enough so that the interactions among them are negligible; meanwhile, by defining an average pressure, density and velocity in the bubbly liquid, the mixture is also assumed to be continuum and the host liquid is incompressible without viscosity so that the following effective equations are valid (Wijngaarden 1968).

$$\frac{D\rho_m}{Dt} + \rho_m \nabla \cdot u = 0$$
$$\rho_m \frac{Du}{Dt} + \nabla p_m = 0 \quad (1)$$
$$p_m = p_m(\rho_m, s)$$

Since we interest in the oscillation at a moderate to high frequency, this system may follow the adiabatic process (Wijngaarden 2007). We then require the thermodynamic behavior of oscillating gas bubbles in the liquid to fulfill an adiabatic process (with constant entropy); thereby the following relations for acoustic waves in water can be applied:

$$\frac{\partial p}{\partial x} = \frac{\partial p}{\partial \rho}\frac{\partial \rho}{\partial x} = c_0^2 \frac{\partial \rho}{\partial x}$$
$$\frac{\partial p}{\partial t} = \frac{\partial p}{\partial \rho}\frac{\partial \rho}{\partial t} = c_0^2 \frac{\partial \rho}{\partial t} \quad (2)$$

The above first-order approximation of the state equation is valid when the nonlinearity of water is negligible. Here $\rho_m = \rho(1-\beta) + \rho_g \beta$ and $\beta = 4\pi n R^3 / 3$ is the volume fraction of bubbles as a function of time and space, $n$ is the bubble number per unit volume, $p_m$ is the average acoustic pressure in the mixture, $c_0$ and $p$ are the sound speed and pressure in the water, respectively.

The above equation can be deduced into a single nonlinear equation (See Appendix for detail derivations)

$$\frac{\partial^2 \rho_m}{\partial t^2} - \frac{\partial^2 p_m}{\partial x^2} = \frac{2}{\rho_m}\left(\frac{\partial \rho_m}{\partial t}\right)^2 + 2u\left(\frac{2}{\rho_m}\frac{\partial \rho_m}{\partial t}\frac{\partial \rho_m}{\partial x} - \frac{\partial^2 \rho_m}{\partial x \partial t}\right) + u^2\left(\frac{2}{\rho_m}\left(\frac{\partial \rho_m}{\partial x}\right)^2 - \frac{\partial^2 \rho_m}{\partial x^2}\right) \quad (3)$$

Eq.(3) is also valid for any other complex fluids by varying the density. By far, compared to the effective equation derived by (Caflisch, Miksis et al. 1985), the conservation of bubble number density has not yet been imposed. Thus, it is also valid to model bubbly liquids in which bubble destructions may happen.

### 2.2. Analysis of the equation

The physical significance of Eq.(3) deduced above is not explicit. In order to obtain some physical meanings for bubbly liquids, we introduce the following parameters to nondimensionalized the above equation

$$p^* = \frac{p_m}{p_0}, \quad x^* = \frac{x}{\lambda} = \frac{x}{c_0/f} = \frac{xf}{c_0}, \quad t^* = tf, \quad \rho^* = \frac{\rho_m}{\rho_0}, \quad u^* = \frac{u\lambda}{fR_0^2} \quad (4)$$

Here the velocity $u$ is scaled relative to the sound speed in bubbly liquids (Caflisch, Miksis et al. 1985). Other reference scaling parameters are atmosphere pressure, wavelength, and frequency of the excitation pressure, as well as the density of water. Substitute the above relations (4) into Eq.(3)

$$\frac{\partial^2 \rho^*}{\partial t^{*2}} - \frac{p_0}{\rho_0 (\lambda f)^2}\frac{\partial^2 p^*}{\partial x^{*2}} = \frac{2}{\rho^*}\left(\frac{\partial \rho^*}{\partial t^*}\right)^2$$
$$+ \frac{2R_0^2}{\lambda^2}u^*\left(\frac{2}{\rho^*}\frac{\partial \rho^*}{\partial t^*}\frac{\partial \rho^*}{\partial x^*} - \frac{\partial^2 \rho^*}{\partial x^* \partial t^*}\right) + \frac{R_0^4}{\lambda^4}u^{*2}\left(\frac{2}{\rho^*}\left(\frac{\partial \rho^*}{\partial x^*}\right)^2 - \frac{\partial^2 \rho^*}{\partial x^{*2}}\right) \quad (5)$$

When $R_0/\lambda \ll 1$, which requires the wavelength to be far larger than the bubble radius, we then have the following simple equation

$$\frac{\partial^2 \rho_m}{\partial t^2} - \frac{\partial^2 p_m}{\partial x^2} = \frac{2}{\rho_m}\left(\frac{\partial \rho_m}{\partial t}\right)^2 \tag{6}$$

which does not require the liquid to be quiescent or stagnant ($u = 0$) though it can be approximated by assuming $u = 0$ in Eq.(5). Eq.(6) is derived without restriction on the pressure, capable of describing the propagation of acoustic waves in other complex fluids as long as the reference length scale is much smaller than the wavelength. Now we can impose the condition for bubbly liquids by substituting $\rho_m \simeq (1-\beta)\rho$ into the above Eq.(6), which requires the bubbly volume fraction is small.

$$(1-\beta)\frac{\partial^2 \rho}{\partial t^2} - \frac{\partial^2 p_m}{\partial x^2} = \rho\frac{\partial^2 \beta}{\partial t^2} - 2\frac{\partial \beta}{\partial t}\frac{\partial \rho}{\partial t} + \frac{2\rho}{1-\beta}\left(\frac{\partial \beta}{\partial t}\right)^2 + \frac{2(1-\beta)}{\rho}\left(\frac{\partial \rho}{\partial t}\right)^2 \tag{7}$$

Use the state relation (2) for the pressure and density, then nondimensionalize Eq.(7) again

$$(1-\beta^*\beta_0)\frac{\partial^2 p^*}{\partial t^{*2}} - \frac{\partial^2 p^*}{\partial x^{*2}} = \frac{\rho_0 c_0^2 \beta_0}{p_0}\rho^*\frac{\partial^2 \beta^*}{\partial t^{*2}} - 2\beta_0\frac{\partial \beta^*}{\partial t^*}\frac{\partial p^*}{\partial t^*}$$
$$+ \frac{2\rho_0 c_0^2 \beta_0^2}{(1-\beta^*\beta_0)p_0}\rho^*\left(\frac{\partial \beta^*}{\partial t^*}\right)^2 + \frac{2(1-\beta^*\beta_0)p_0}{\rho_0 c_0^2}\frac{1}{\rho^*}\left(\frac{\partial p^*}{\partial t^*}\right)^2 \tag{8}$$

where $\beta_0 = 4\pi N_0 R_0^3/3$, $N_0$ is the initial bubble numbers per unit volume, $R_0$ is the initial bubble radius.

The Eq.(7) and Eq.(8) are second order PDEs that contain nonlinear terms of the propagating finite amplitude pressure waves of arbitrary order and volume fraction of lowest order. The nondimensional form can be further reduced by simple mathematical analysis. For instance, the condition for neglecting nonlinear terms require that

$$\frac{2\rho_0 c_0^2 \beta_0^2}{(1-\beta^*\beta_0)p_0} \ll 1, \quad \frac{2(1-\beta^*\beta_0)p_0}{\rho_0 c_0^2} \ll 1 \tag{9}$$

For the case we are studying, it is evident that $\rho_0 c_0^2 \sim 10^9$ and $p_0 \sim 10^5$. In the present model without considering multiple scattering and bubble-bubble interactions, $\beta_0 \ll 10^{-4}$. the two conditions in (9) are satisfied automatically, thus we have

$$\frac{1}{c_0^2}\frac{\partial^2 p}{\partial t^2} - \frac{\partial^2 p_m}{\partial x^2} = \rho\frac{\partial^2 \beta}{\partial t^2} \tag{10}$$

Here the pressure $p_m$ (in the mixture) has not yet been determined. Repeating the derivation of state equation of bubbly liquids $\partial p_m / \partial x = c^2 \partial \rho_m / \partial x$, for the lowest order approximation of the volume fraction, it is not difficult to obtain $\partial^2 p_m / \partial x^2 = c^2 \partial^2 \rho / \partial x^2$. Here $c$ is the sound speed in the bubbly liquids. Using the first relation in Eq.(2), we have

$$\frac{1}{c_0^2}\frac{\partial^2 p}{\partial t^2} - \frac{c^2}{c_0^2}\frac{\partial^2 p}{\partial x^2} = \rho_0 \frac{\partial^2 \beta}{\partial t^2} \qquad (11)$$

which is the major result proposed in this study. The sound speed $c$ in the bubbly liquids can be determined using (Xia 2018)

$$c^2 = c_0^2 + \frac{B}{A} c_0^2 \frac{\rho_m'}{\rho_0} + O(3) \qquad (12)$$

Since we have assumed the bubble volume fraction to be small, then the above equation can be reduced to

$$c^2 \approx c_0^2 (1 - \frac{B}{A}\beta_0) \qquad (13)$$

Thus Eq.(11) becomes

$$\frac{1}{c_0^2}\frac{\partial^2 p}{\partial t^2} - (1-\frac{B}{A}\beta_0)\frac{\partial^2 p}{\partial x^2} = \rho_0 \frac{\partial^2 \beta}{\partial t^2} \qquad (14)$$

.

Eq.(14) is the main result of this paper.

### 3. Discussion

*Classical results*

Recall the wave equation in bubbly liquids (Caflisch, Miksis et al. 1985, Commander and Prosperetti 1989),

$$\frac{1}{c_0^2}\frac{\partial^2 p}{\partial t^2} - \frac{\partial^2 p}{\partial x^2} = \rho_0 \frac{\partial^2 \beta}{\partial t^2} \qquad (15)$$

it can be treated as acoustic waves propagating in water with source term due to bubble oscillations added to the right of the equation. Thus the nonlinearity in the bubbly liquids is caused by the nonlinear oscillations of the bubbles and their interactions with the acoustic waves. The present equation demonstrates how the impact of dispersion (related by $c$) on the nonlinear propagation

From the derivation above we can see that the Caflisch's equation (15) is the first order approximation of Eq.(7) with also imposing

$$\frac{B}{A}\beta_0 \ll 1 \qquad (16)$$

This condition is valid as long as the frequency of the propagating wave is far away from the resonance frequency of the bubbles in the bubbly liquid. Here B/A is the nonlinearity parameter of the bubbly liquid depending on the volume fraction $\beta_0$ and wave frequency (Xia and Sarkar 2017, Xia 2018). Although more complicated cases can be made by the analysis of the pressure $p_m$ in the mixture, we leave it for future exploitation.

*Anderson localization*

If the bubbles in the medium are assumed to be monodisperse, and the volume fraction to be $10^{-5}$, the nonlinearity parameter could be $10^5$ when bubbles oscillate close to their resonance frequency (Xia and Sarkar 2017). Thus the spatial term in Eq.(14) could disappear, indicating the pressure waves is not able to propagate in the medium. Localization of the acoustic wave could happen in bubbly liquids when the bubbles oscillate at resonance. This result is significant also when the bubbly liquid is treated as an acoustic metamaterial because one can control the propagation of the acoustic waves by the manipulate the concentration of the bubbles, as well as the resonance frequency.

Future studies can focus on a second order approximation of the average pressure in the mixture based on the present model, as well as the thermal effects at resonance. The derivation of second-order nonlinear waves equations in bubbly liquids can also be done to check the accuracy of the first order approximation of volume fraction.

**Appendix**

Derivation of the nonlinear wave equation Eq.(3) for waves propagation in bubbly liquids is as follows:

Following the methodology of Wijngaarden-Papanicolaou model, we may begin with basic conservation laws for one-dimensional plane waves that widely used in describing bubbly flow:

1. Conservation of mass:

$$\frac{D}{Dt}\rho_m + \rho_m \frac{\partial u}{\partial x} = 0$$

2. Conservation of momentum:

$$\rho_m \frac{D}{Dt} u + \frac{\partial p_m}{\partial x} = 0$$

where the material derivative

$$\frac{D}{Dt} = \frac{\partial}{\partial t} + u \frac{\partial}{\partial x}$$

and $\rho_m = (1-\beta)\rho + \beta \rho_g$. Here $\rho_m$ is the average local mixture density, $\rho$ and $\rho_g$ are the host liquid and gas density respectively. The local fraction of volume occupied by the gas is given by $\beta = 4\pi n R^3 / 3$, where $R$ is the instantaneous radius of the bubbles and $n$ is their number per unit volume.

Reorganize the above equations of conservations

$$\begin{cases} \dfrac{\partial u}{\partial x} = -\dfrac{1}{\rho_m} \dfrac{D}{Dt} \rho_m \\[1em] \dfrac{D}{Dt} u = -\dfrac{1}{\rho_m} \dfrac{\partial p_m}{\partial x} \end{cases}$$

$$\begin{cases} \dfrac{D}{Dt}\left(\dfrac{\partial u}{\partial x}\right) = -\dfrac{D}{Dt}\left(\dfrac{1}{\rho_m} \dfrac{D}{Dt} \rho_m\right) \\[1em] \dfrac{\partial}{\partial x}\left(\dfrac{D}{Dt} u\right) = -\dfrac{\partial}{\partial x}\left(\dfrac{1}{\rho_m} \dfrac{\partial p_m}{\partial x}\right) \end{cases}$$

Note also that

$$\frac{\partial}{\partial x}\left(\frac{D}{Dt} u\right) = \frac{D}{Dt}\left(\frac{\partial u}{\partial x}\right) + \left(\frac{\partial u}{\partial x}\right)^2$$

Use the equation of conservation of mass again

$$\frac{\partial}{\partial x}\left(\frac{D}{Dt} u\right) = \frac{D}{Dt}\left(\frac{\partial u}{\partial x}\right) + \left(-\frac{1}{\rho_m} \frac{D}{Dt} \rho_m\right)^2$$

thus

$$\left(-\frac{1}{\rho_m}\frac{D}{Dt}\rho_m\right)^2 - \frac{D}{Dt}\left(\frac{1}{\rho_m}\frac{D}{Dt}\rho_m\right) = -\frac{\partial}{\partial x}\left(\frac{1}{\rho_m}\frac{\partial p_m}{\partial x}\right)$$

Expand the above equation

$$\frac{2}{\rho_m}\left(\frac{D}{Dt}\rho_m\right)^2 - \frac{D}{Dt}\left(\frac{D}{Dt}\rho_m\right) = \frac{1}{\rho_m}\frac{\partial \rho_m}{\partial x}\frac{\partial p_m}{\partial x} - \frac{\partial^2 p_m}{\partial x^2}$$

Remember that

$$\frac{D}{Dt}\left(\frac{D}{Dt}\rho_m\right) = \frac{D}{Dt}\left(\frac{\partial \rho_m}{\partial t} + u\frac{\partial \rho_m}{\partial x}\right) = \left(\frac{\partial}{\partial t} + u\frac{\partial}{\partial x}\right)\left(\frac{\partial \rho_m}{\partial t} + u\frac{\partial \rho_m}{\partial x}\right)$$

$$= \frac{\partial^2 \rho_m}{\partial t^2} + 2u\frac{\partial^2 \rho_m}{\partial x \partial t} + u^2\frac{\partial^2 \rho_m}{\partial x^2} + \left(\frac{D}{Dt}u\right)\left(\frac{\partial \rho_m}{\partial x}\right)$$

Take material derivatives to the momentum equation

$$\frac{D}{Dt}\left(\frac{D}{Dt}\rho_m\right) = \frac{\partial^2 \rho_m}{\partial t^2} + 2u\frac{\partial^2 \rho_m}{\partial x \partial t} + u^2\frac{\partial^2 \rho_m}{\partial x^2} - \frac{1}{\rho_m}\left(\frac{\partial p_m}{\partial x}\right)\left(\frac{\partial \rho_m}{\partial x}\right)$$

then

$$\frac{2}{\rho_m}\left(\frac{\partial \rho_m}{\partial t} + u\frac{\partial \rho_m}{\partial x}\right)^2 - \left(\frac{\partial^2 \rho_m}{\partial t^2} + 2u\frac{\partial^2 \rho_m}{\partial x \partial t} + u^2\frac{\partial^2 \rho_m}{\partial x^2} - \frac{1}{\rho_m}\left(\frac{\partial p_m}{\partial x}\right)\left(\frac{\partial \rho_m}{\partial x}\right)\right)$$

$$= \frac{1}{\rho_m}\frac{\partial \rho_m}{\partial x}\frac{\partial p_m}{\partial x} - \frac{\partial^2 p_m}{\partial x^2}$$

Rearrange the above equation

$$\frac{\partial^2 \rho_m}{\partial t^2} - \frac{\partial^2 p_m}{\partial x^2} = \frac{2}{\rho_m}\left(\frac{\partial \rho_m}{\partial t}\right)^2 + 2u\left(\frac{2}{\rho_m}\frac{\partial \rho_m}{\partial t}\frac{\partial \rho_m}{\partial x} - \frac{\partial^2 \rho_m}{\partial x \partial t}\right) + u^2\left(\frac{2}{\rho_m}\left(\frac{\partial \rho_m}{\partial x}\right)^2 - \frac{\partial^2 \rho_m}{\partial x^2}\right)$$

The derivation of above equation employed the characteristics of the One-dimensional form of the PDEs; for higher dimensions, however, the reduction is not valid.